\documentclass{article}
\usepackage{epsf}
\usepackage{emulateapj,pstricks}
\input{psfig}
\usepackage{mymathptm}
\usepackage{flushrt}

\lefthead{}

\def\mathfont#1{\ifmmode{#1}\else{$#1$}\fi} 
\def\lae{\mathrel{<\kern-1.0em\lower0.9ex\hbox{$\sim$}}}  
\def\gae{\mathrel{>\kern-1.0em\lower0.9ex\hbox{$\sim$}}}  
\def\kms{\ifmmode{{\rm km\ s}^{-1}}\else{${\rm km\ s}^{-1}$}\fi} 

\def\ergsec{\mathfont{ {\rm ergs\ s}^{-1}}}

\def\msun{\ifmmode{\ {\rm M}_\odot}\else{$ {\rm M}_\odot$}\fi}  
\def\msunyr{\ifmmode{\msun \ {\rm yr}^{-1}}\else{$\msun \ {\rm yr}^{-1}$}\fi}

\righthead{CHANDRA X-RAY OBSERVATIONS OF THE HYDRA A CLUSTER: AN INTERACTION BETWEEN THE RADIO SOURCE AND THE X-RAY-EMITTING GAS}

\slugcomment{submitted to {\em The Astrophysical Journal Letters}}
\def\ref#1{\noindent\hangindent=24.0pt\hangafter=1{#1}\par}
\def\la{\hbox{\rlap{$<$}\lower.5ex\hbox{$\sim$}\ }}
\def\ga{\hbox{\rlap{$>$}\lower.5ex\hbox{$\sim$}\ }}

\begin{document}

\lefthead{}
\righthead{Chandra X-ray Observations of Hydra A}

\slugcomment{submitted to {\em The Astrophysical Journal Letters}}

\title{CHANDRA X-RAY OBSERVATIONS OF THE HYDRA A CLUSTER: AN
INTERACTION BETWEEN THE RADIO SOURCE AND THE X-RAY-EMITTING GAS}

\author{B.~R.~ McNamara,\altaffilmark{1} 
M. ~Wise,\altaffilmark{3}
P.~E.~J.~Nulsen,\altaffilmark{1,2} 
L.~P.~David,\altaffilmark{1} 
C.~L.~ Sarazin,\altaffilmark{4}
M.~ Bautz,\altaffilmark{3}
M.~Markevitch,\altaffilmark{1}
A. Vikhlinin,\altaffilmark{1}
W.~R.~Forman,\altaffilmark{1}
C.~Jones,\altaffilmark{1}
D.~ E.~ Harris,\altaffilmark{1}}

\bigskip

\doublespace

\begin{abstract}

We present Chandra X-ray Observations of the Hydra A 
cluster of galaxies, and we report the discovery of structure in
the central 80 kpc of the cluster's X-ray-emitting gas.
The most remarkable structures are depressions in the X-ray surface 
brightness, $\sim 25-35$ kpc diameter, that are
coincident with Hydra A's radio lobes.
The depressions are nearly devoid of X-ray-emitting gas,
and there is no evidence for shock-heated gas surrounding the radio
lobes.  We suggest the gas within the surface brightness
depressions was displaced as the radio lobes expanded subsonically, 
leaving cavities in the hot atmosphere. 
The gas temperature declines from 4 keV at 70 kpc to 3 keV in the inner 20 kpc
of the brightest cluster galaxy (BCG), and the
cooling time of the gas is $\sim 600$ Myr in the inner 10 kpc.  These
properties are consistent with the presence of a $\sim 34 \msunyr$
cooling flow within a 70 kpc radius.  
Bright X-ray emission is present in the BCG
surrounding a recently-accreted disk of nebular emission and young stars.
The star formation rate is commensurate with the cooling rate 
of the hot gas within the volume of the disk, although the
sink for the material cooling at larger radii remains elusive. 
A bright, unresolved X-ray source is present in the BCG's nucleus,
coincident with the radio core.  It's X-ray spectrum 
is consistent with a power law absorbed by a foreground  
$N_{\rm H}\simeq 4\times 10^{22}~{\rm cm}^{-2}$ column of hydrogen.  This
column is roughly consistent with the hydrogen column seen in absorption
toward the $\lae 24$ pc diameter VLBA radio source.  Apart from the
point source, no evidence for excess X-ray absorption above 
the Galactic column is found.
\end{abstract}

{\it Subject headings:}  Galaxies: clusters: individual (Hydra A) -- cooling
flows -- intracluster medium 
\vfill

\section{Introduction}
The Hydra A radio galaxy is associated with a relatively poor
cluster of galaxies at redshift $z=0.052$.  The cluster harbors an atmosphere
of X-ray emitting gas of luminosity
$L_x(0.5-4.5)=2.2\times 10^{44}~\ergsec$, and
a mean gas temperature of $\sim 4$ keV based on
$Einstein$ MPC observations (David et al.\ 1990) and ASCA observations
(Ikebe et al.\ 1997).
%
%
A cooling flow is present with an 
accretion rate of ${{\dot M}}\sim 250 \msunyr$ (assuming
${\rm H_0}=50~{\rm km ~s^{-1}~Mpc^{-1}}$) within a radius
of $\simeq 170$ kpc (White et al.\ 1997; Peres et al.\ 1998).
The hot atmosphere is centered on
a brightest cluster elliptical galaxy (BCG) that 
hosts a large (80 arcsec or 84 kpc in projection), unusually powerful
($P =1.6\times 10^{26}~ {\rm W~ Hz^{-1}}$ at 178 MHz)
Fanaroff-Riley type 1 (FR 1) radio source 3C 218 
(Ekers \& Simkin 1983; Taylor et al.\ 1990; Taylor 1996).  
The twin jet-lobe radio source emerges from
a disk of young stars (McNamara 1995; Hansen et al.\ 1995;
Melnick et al. 1997) and nebular emission (Ekers \& Simkin 1983;
Hansen et al.\ 1995; Baum et al.\ 1988) several kpc
in size that is in rotation about the
nucleus.  H I is seen in absorption toward the nucleus (Dwarakanath, Owen, \& van Gorkom 1995) and has been mapped in absorption
against the parsec scale nuclear radio source with the VLBA
(Taylor 1996).  

Hydra A was observed by the Chandra X-ray Observatory
during its orbital verification and activation phase.
Hydra A's several interesting properties provide 
a first opportunity to investigate a cooling flow and potential
interactions between

\footnotesize
$^1$Harvard-Smithsonian Center for Astrophysics, 60 Garden St. Cambridge, MA 02138

$^2$Department of Engineering Physics, University of Wollongong, Wollongong NSW 2522, Australia

$^3$Massachusetts Institute of Technology, Center for Space Research, Cambridge, MA 02139

$^4$Astronomy Department, University of Virginia, Charlottesville, VA


\normalsize
\noindent
the radio 
source and X-ray-emitting gas using Chandra's unprecedentedly high 
spatial resolution. We present an analysis of Hydra A based on
the preliminary telescope calibration in this {\it Letter},
and report the discovery of
an interaction between the radio source and the X-ray-emitting
gas.  We assume ${\rm H_0}=70~{\rm km ~s^{-1}~Mpc^{-1}}$,
$\Omega_{\rm M}=0.3$, $\Omega_\Lambda =0.7$, a luminosity distance of 240 Mpc,
and 1 arcsec = 1.05 kpc.

\section{Data Analysis}

The calibration observations of Hydra A were performed on 1999 October
30.  A total integration time of 40 ksec was obtained, 
20 ksec centered at the aim point
of the S3 back-illuminated ACIS chip (OBSID 576), and 20 ksec at the ACIS I
aim point on the I3 front-illuminated device (OBSID 575).  X-ray events with
energies below 300 eV and above 10 keV were not considered in
our analysis, and flight grades 0, 2, 3, 4, 6 were retained.
The particle background was generally stable throughout the
S3 observations.  Only 300s of the 20 ksec exposure
experienced a $30\%$ increase in particle background.  Therefore
no S3 data were rejected on this basis.  The spectral analysis
presented here is limited to a $0.5 - 7$ keV bandpass for the S3 data. 


\subsection{X-ray Morphology}

A Chandra image of the X-ray emission from the central $118\times 118$
arcsec of the cluster is shown in Fig. 1.  The image is a
summed, 40 ksec exposure obtained with the ACIS-I3, front
illuminated device, and the ACIS-S3, back-illuminated device.
The top panel shows the unsmoothed image, after event filtering,
centered on the X-ray point source. 
The X-ray emission shows a great deal of structure
on scales ranging from less than a few kpc to tens of kpc that
have not been seen in earlier X-ray imagery.  An X-ray point
source ($RA=09~18~05.77$ , $Dec=-12~05~42.53$, J2000), 
shown inset to the top panel of
Fig. 1,  coincides with the central radio
core and BCG nucleus.  The lower panel of Fig. 1 shows
the wavelet-smoothed X-ray image.  The emission
is concentrated in a central triangular region, 10--15 arcsec in
size, and in fainter fingers of emission extending northeast
and southwest of the center.  The emission spectrum of this
material is consistent with thermal emission from $\sim 3$ keV
cluster gas.  Two depressions in the X-ray emission, 20--30 arcsec
in diameter, are seen $\sim 20$ arcsec to the north and south of the cluster
center. 
The surface brightness within these cavities is a factor of
$\sim 1.5$ lower than the mean surface brightness of the emission
at similar radii from the center.  This decline in surface brightness
is consistent with these regions being devoid of gas
at the ambient temperature and density.

\begin{center}

\psfig{file=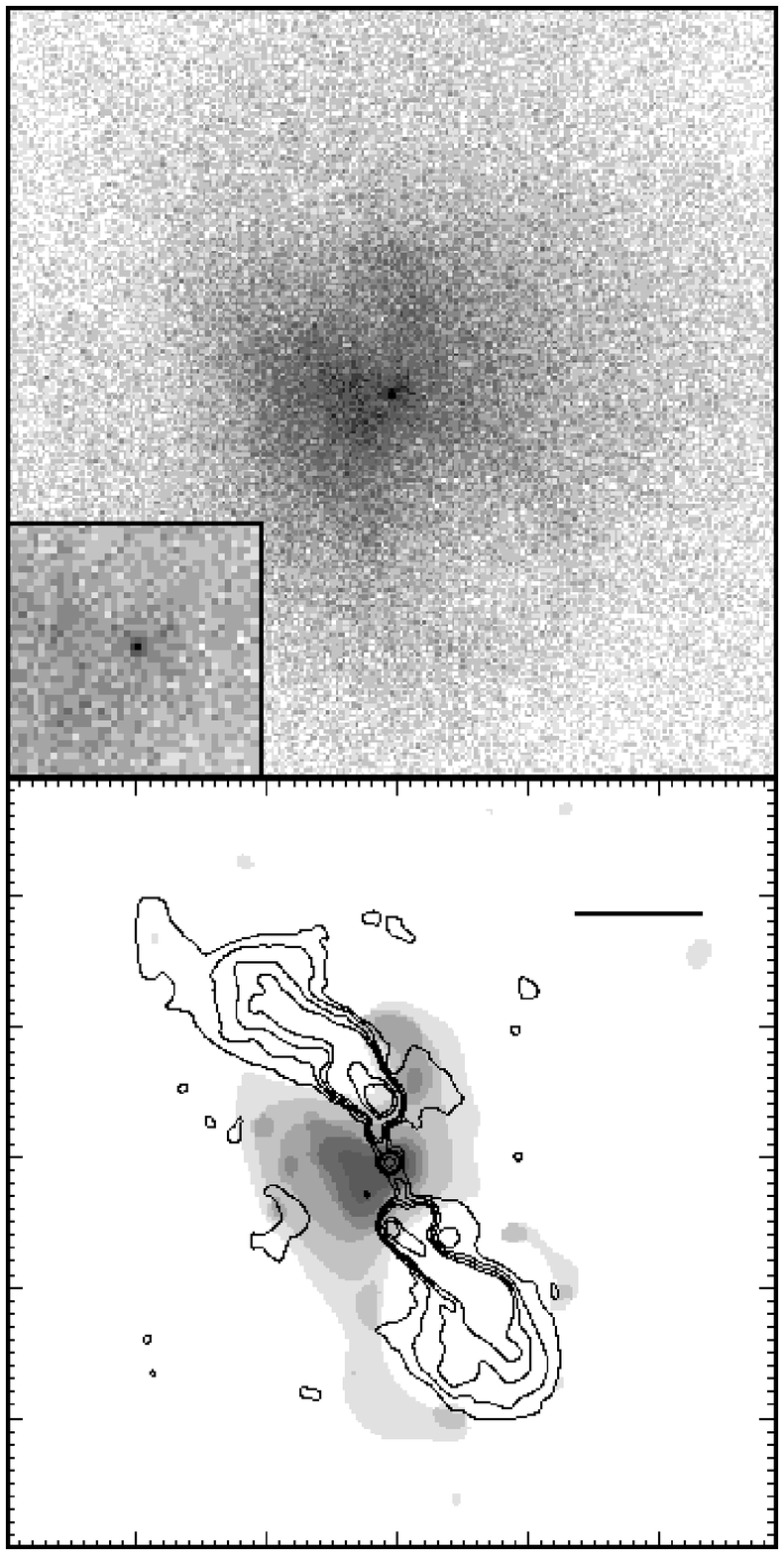,height=14.0cm,width=7.0cm}
\begin{minipage}{8cm}
\smallskip
\small\parindent=3.5mm {\sc Fig.}~1. 
{\it Top Panel:} 40 ksec integration X-ray
image of the central region of the Hydra A cluster centered on the
nuclear X-ray point source (shown inset). {\it Bottom Panel:} Wavelet
smoothed and reconstructed image of the same region superposed on
the 6 cm VLA Radio image of Hydra A
(contours). The scale bar is 20 arcsec in length.     
\end{minipage}

\end{center}







\subsection{Radio Morphology}

A 6 cm VLA image obtained in the A-array, obtained by G. Taylor,
is shown superposed on the wavelet-reconstructed
X-ray image in Fig. 1.  The cavities in
the X-ray emission are filled by emission from the
radio lobes, and the X-ray cavities and radio lobes are
remarkably similar in shape.
Similar cavities have been seen in $ROSAT$ imagery of the Perseus
cluster (B\"ohringer et al.\ 1992) and in Cygnus A (Carilli et al.\ 1994).  Heinz et al.\ (1998) proposed a
model to explain such cavities as a shell of shocked gas 
displaced by the relativistic material from the radio jet.
This model would imply that the X-ray emission
surrounding the radio source should be considerably hotter than the
material away from the radio source.  We
evaluate this interpretation of the radio cavities, and we explore 
the physical state of the cluster gas in \S 2.4.

\subsection{Optical Morphology: The Central Disk}

Spatial correlations between the X-ray, optical,
and radio emission are present in the inner 20 kpc
of the BCG.  The $U$-band contours are superposed on a 
grayscale X-ray image in Fig. 2.
The X-ray point source is centered on a gaseous disk
of nebular emission and young blue stars in rotation
about the nucleus (Ekers \& Simkin 1983; Baum et al.\ 1988;
McNamara 1995; Hansen et al.\ 1995; Melnick et al.\ 1997).
The disk of star formation is about 10 kpc by 7 kpc in size,
and is roughly $0.7$ magnitudes bluer than the $U-I$ 
color of the galaxy's halo (McNamara 1995).
The nebular emission (Baum et al.\ 1988; Hansen et al.\ 1995) 
extends over a similar region.

The brightest X-ray emission, apart from the point source,
appears in a flattened structure 
coincident with the disk, and in an irregular structure
several arcsec east of the disk.  The X-ray point source is
within a half arcsec of the radio core (Fig. 1) and the BCG's nucleus.  
This positional correspondence is within Chandra's absolute pointing
error. The surface brightness
of the light gray region in Fig. 2 is $\simeq 30\%$
fainter than the dark gray structure just east of the disk. 
This fainter emission forms a wedge-shaped structure that widens 
toward two companion galaxies projected onto the BCG (see also Fig. 1).
The asymmetry in the X-ray structure may be caused, in part, by an irregular
potential well associated with a merger between the BCG and
the companion galaxies (Ekers \& Simkin 1983).

\begin{center}
\psfig{file=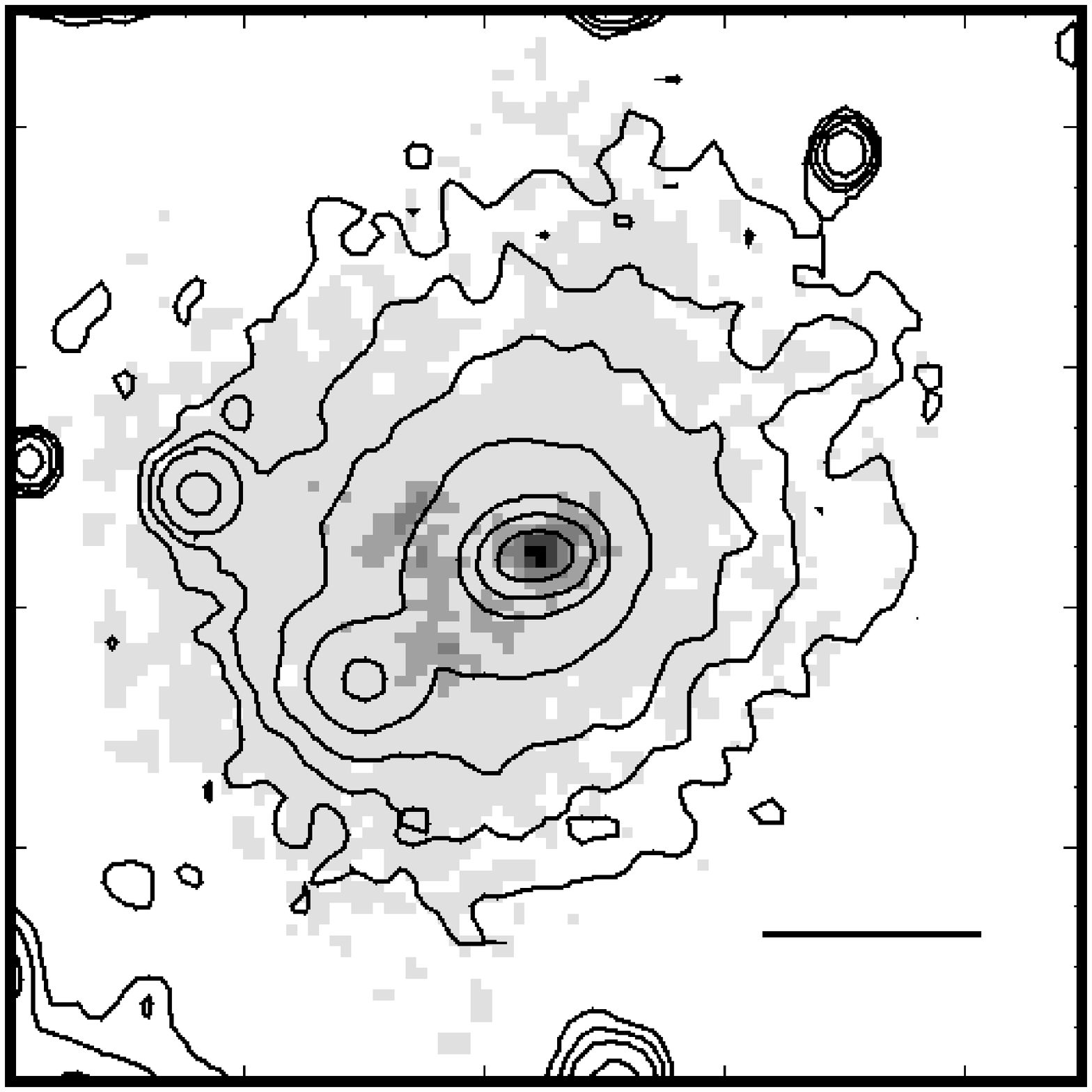,height=7.0cm,width=7.0cm}
\begin{minipage}{8cm}
\smallskip
\small\parindent=3.5mm {\sc Fig.}~2. Detail of the X-ray emission,
grayscale smoothed with a $FWHM=1.75$ pix gaussian, 
superposed on the $U$-band contours of the BCG (McNamara 1995).  
The central disk of gas and young stars are the
elliptical contours centered on the X-ray point source. 
The scale bar is 10 arcsec in length.
\end{minipage}
\end{center}

The radio jets (Fig. 1) emerge from the optical disk at roughly a $20\deg$
angle from the disk's minor optical axis and rotation axis, and
flare into lobes where the gas pressure reaches 
$\sim 6\times 10^{-10}~{\rm erg~cm^{-3}}$.
The jets show no obvious signs of interacting with the
optical disk emission or the X-ray emission in the central flattened
structure.  


\subsection{Physical State of the X-ray Emitting Gas}

We computed the radial distribution
of density, temperature, and pressure of the gas in the central
84 kpc.  These parameters were computed by fitting a single
temperature XSPEC model to the X-ray emission in
annuli centered on a central X-ray point source. The fluxes were
corrected for cluster emission at large radii projected onto the inner
regions (deprojected), and the abundance and foreground absorption were free
parameters in the models. The abundances were found to be $\sim 0.4$
solar, and the foreground absorption $\sim 2 \times 10^{20}~{\rm cm}^{-2}$. 
No evidence for excess absorption from cold gas
within the cluster is found, with the exception of a large
column toward the nuclear point source (\S 4).  In addition, 
we constructed hardness ratio profiles, $\kappa(R)$, by taking the ratios
of the X-ray surface brightnesses, $I$, in several passbands.  
These profiles are plotted in Fig. 3.
The hardness ratios are defined as  
$\kappa_1=I(0.5\rightarrow 1.5)/I(1.5\rightarrow 2.5)$ (open symbols) 
and $\kappa_2=I(0.5\rightarrow 2.5)/I(2.5\rightarrow 6.0)$ (filled
symbols), where the figures in brackets are the passbands in keV.
The spectrum is harder as $\kappa (R)$ decreases. 
The $n_e$, $kT$, and $P$ profiles
exclude the central point source, while the hard (absorbed) 
point source is included in the hardness ratio plot.  

The temperature increases from $\sim 3$ keV in the central
10 kpc to $\sim 4$ keV at 80 kpc.  The rms density within the central 
10 kpc is $n_e=0.06~ {\rm cm}^{-3}$.
The density declines with radius as $\sim r^{-1}$ to a radius of 70
kpc, and 
the gas pressure declines by a factor of $\sim 6$ from the center to 70 kpc.  
The central point source is considerably harder than the
surrounding emission within a 10 kpc radius.
Beyond the central source, $\kappa$ decreases (hardens) $10\%-30\%$ between the center and 80
kpc. The broad-band $\kappa_2$ shows the larger decline.  
There is no
significant difference between hardness ratio profiles 
including or excluding the flux from the cavities.

We note that in the 0.5 -- 7 keV Chandra band, the count rate for gas
at 80 keV is about 2/3 of that for gas with the same emission measure
at 5 keV.  If gas in or near the cavities has been compressed in a
shock, its emission in the Chandra band would be larger than from the
same amount of gas at ambient conditions.  Then the regions within
the radio lobes would be brighter than the surrounding emission,
which is not observed.

\section{Discussion}

\subsection{Origin of the X-ray Cavities}

There is no indication that the gas surrounding the radio source is hotter than
the ambient cluster gas.  
This behavior is inconsistent with strongly shock heated
gas, as is suggested by the Heinz et al.\ (1998) model, 
but is consistent with the cool central material
being displaced as the radio source expands subsonically.
There is no evidence that the radio source is heating the gas outside
the cavities (Nulsen et al.\ in preparation). 

\begin{center}
\psfig{file=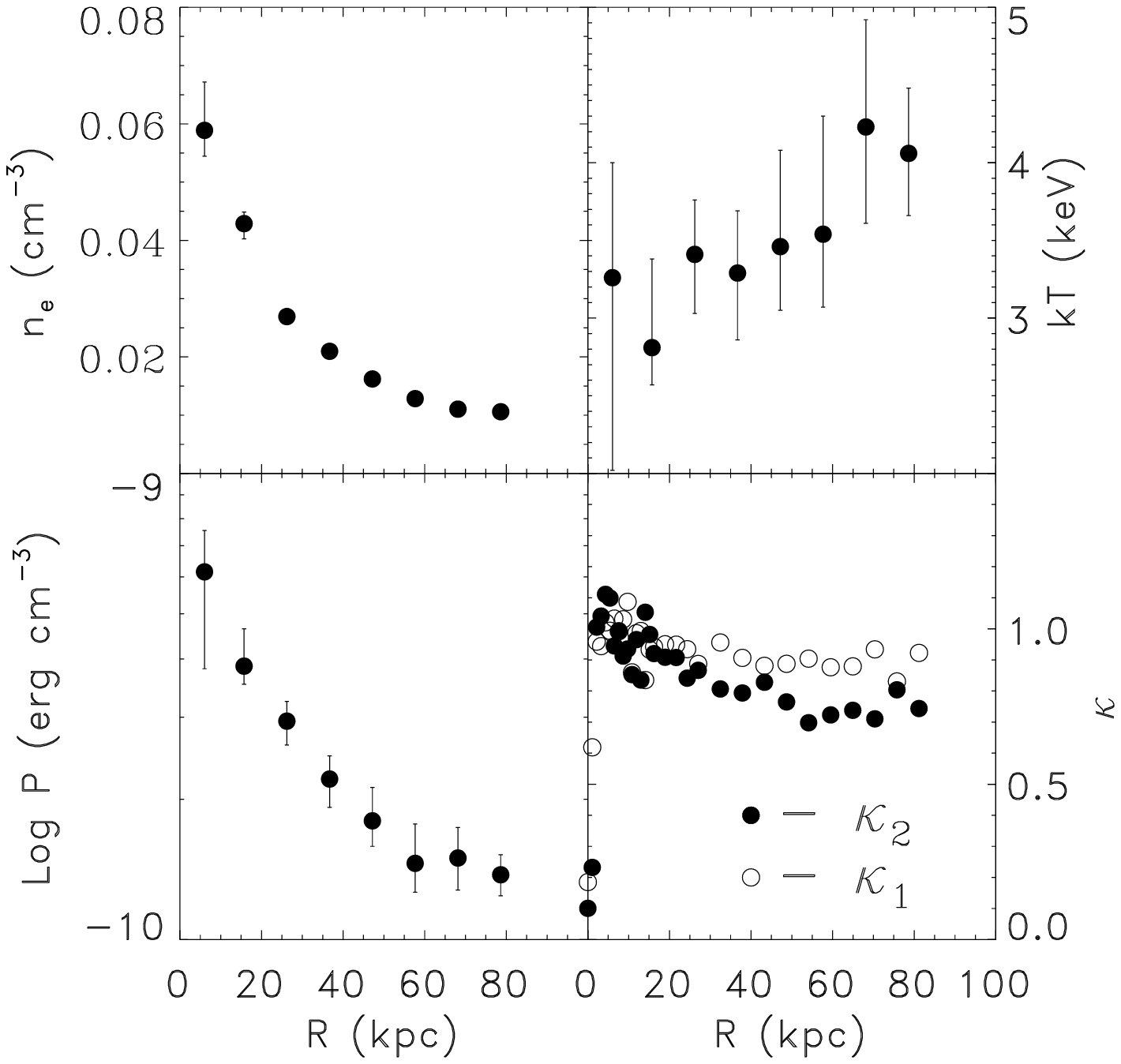,height=7.9cm,width=7.9cm}
\begin{minipage}{8cm}
\smallskip
\small\parindent=3.5mm {\sc Fig.}~3. Run of electron density (upper left),
temperature (upper right), pressure (lower left), and hardness ratio
(lower right).  In this convention, hardness increases as $\kappa$ decreases.

\end{minipage}

\end{center}

We now calculate the minimum energy required by the
radio source to displace the gas using the southern cavity
for illustration.  The center of the southern
cavity is approximately $26$ arcsec from the X-ray nuclear point
source. The
electron density in the annulus centered on the southern cavity
is $n_{\rm e} = 0.027 \rm \, cm^{-3}$ and the
temperature is $kT = 3.4 \rm \,~ keV$, giving a pressure $p = 2.8
\times 10^{-10} \rm \, erg \, ~cm^{-3}$.
Based on the deprojection, undisturbed gas in a volume the size of
this cavity would have contributed about 2000 counts. This is
consistent with our estimate for the count deficit in the cavity.  We
therefore assume that the ambient ICM has been displaced from this
region by the radio source.  Since there is no spectral
hardening at the edges of the cavities, we assume that the pressure
within them is similar to the ambient gas pressure.

The minimum energy required to push the gas
out of a sphere of radius 15 kpc is then $ p V \simeq 1.2 \times 10^{59}
\rm \, erg$.  Since there are no signs of shocks, we assume
that gas motions are subsonic and can be
ignored, so that the total energy required to inflate the cavity is a
modest multiple of this.

The cavity would take $\sim 2\times 10^7$ yr to expand at 
approximately the sound speed.  The cavity is also 
buoyant, so it must be refilled on a similar timescale,
$\simeq 2 R\sqrt{r / GM(R)}$, where $r$ is the radius of
the cavity and $R$ is the distance to the cluster center.  
From the deprojection, a rough estimate of the
total gravitating mass within 30 kpc of the cluster center is
$M(R)=3\times10^{12} \rm \, M_\odot$, giving a refilling time of about
$6\times10^{7} \rm \, yr$.
Thus, the minimum mechanical power from the southern radio jet required
to maintain the cavity is $\sim 6\times10^{43} \rm \, erg \, s^{-1}$,
which is comparable to the total radio power from 
Hydra A (Ekers \& Simkin 1983).  
Either timescale implies an efficiency of conversion of kinetic energy to radio
power of order unity, and we need not postulate the existence of
kinetic luminosity substantially in excess of the luminous radio power 
(e.g. Heinz et al.\ 1998).  

Our interpretation requires
the radio lobes and surrounding gas to be nearly in pressure equilibrium.
However, the gas pressure appears to be  more than
an order of magnitude larger than the minimum energy pressure
of the radio lobes (Taylor et al.\ 1990).  
If the lobes are in pressure balance with the cluster gas, and if 
equipartition between the magnetic field and relativistic particles is to
be maintained, then one or more of the following conditions 
is implied. There is a significant additional contribution to the particle 
energy density from low energy electrons or from protons
(e.g. B\"ohringer et al.\ 1993), the radio 
filling factior is significantly less than one, or the radio lobes are
at significantly larger radii in the cluster atmosphere 
than their projected positions.

\subsection{The Cooling Flow}

Gas with the observed central density 
$n_e(r<10~{\rm kpc})=0.06~ {\rm cm}^{-3}$,
temperature $kT=3$ keV, and abundance of $0.4$ solar
has a radiative cooling time of $\sim 6\times 10^8$ yr. 
If such cool material has been present in the cluster
for $\gae$1 Gyr, the gas should cool to low
temperatures and flow to the center of the cluster.
The Chandra data allow us to obtain high
quality spectra of several independent
regions within the central 80 kpc of the cluster.
%
%
We have fitted an absorbed thermal (mekal) plus constant
pressure cooling flow (mkcflow) model to a
series of 8 circular apertures ranging in equal steps from $10$ arcsec to $80$
arcsec in radius.  The central $1.5$ arcsec was excluded.  
Assuming spherical symmetry, we attribute the differences
in mass deposition rate, ${{\dot M}}$, 
for successive apertures to the annulus
they define, apply a standard geometric deprojection to convert
these to mass deposition rates per unit volume,
and calculate total ${{\dot M}}$'s for the spheres.  
The total mass deposition rate within a 74
kpc sphere is $34 \pm 5 \msunyr$, and ${{\dot M}}(R)\sim R^{1.1}$.

\subsection{Cooling Rate vs Star Formation Rate}

The most controversial issue concerning cooling flows
is the fate of the cooling material.  We now ask whether
the accretion
rates derived from the new X-ray data are consistent with the 
star formation rate.
Star formation appears to be confined to the disk (\S2.3), whose
young stellar population mass estimate ranges from
$10^{7.7\rightarrow 9}\msun$ (McNamara 1995; Hansen et al.\ 1995).  
Estimates of the average star formation rates vary from 
$\lae 1 \msunyr$ for continuous
star formation for $\sim$Gyr duration, to $\sim 15 \msunyr$
for a younger, $\sim 10^8$ yr old burst population.  The degree to which star
formation may be consuming the cooling gas
depends on factors including the relative ages of the cooling flow and the 
stellar population, and the stellar initial mass function.
In the central 10
kpc, a region somewhat larger than the central optical disk, 
the mass deposition rate is $4 \pm 2\msunyr$.
Thus, the observed rates of star formation can account entirely 
for the mass cooling within this region regardless of the age of
the cooling flow. However, the disk star formation 
would provide a sink for the cooling gas within the entire 
74 kpc region for $\lae 3\times 10^6$ yr, which is much
smaller than the minimum time $\sim 1$ Gyr required to 
establish a cooling flow.  Most of the cooling material remains
unaccounted for, and there is no evidence that the radio source is
heating the gas and reducing the cooling rates.

\section{The Central Point Source}

We extracted the spectrum of the central point source within a
1 arcsec radius aperture. The background was taken from the
annular region between $r=1-2.5$ arcsec centered
on the point source.  The net point source spectrum was 
found by subtracting the background spectrum, after normalizing by
the relative areas of the central and annular regions.
The net spectrum appears to be a power law absorbed by 
a $N_{\rm H}= 4.5 (2.2-9.9)\times 10^{22}~{\rm cm}^{-2}$ ($90\%$ errors).  
The point source flux, uncorrected for absorption, is
$f(0.6-6.0)=1.0\times 10^{-13}~ \ergsec~{\rm cm}^{-2}$, which
corresponds to a luminosity of $6.9\times 10^{41}~\ergsec$.
The absorbing column toward the point source is
more than two orders of magnitude larger than the Galactic foreground column,
and is confined to a spatial extent of $< 1.5$ arcsec.

This column density is in reasonably good agreement with the column of neutral
hydrogen absorption seen toward the VLBI radio source
in the nucleus of the galaxy (Taylor 1996).  The radio absorption
map confines the spatial extent of this high column density material
to a region of $\lae 24$ pc  in the nucleus of the galaxy.
Assuming the X-ray and radio flux is absorbed by the same material,
the combined observations restrict the X-ray-emitting region
of the point source to $\lae 24$ pc.

\medskip

\noindent

\section*{Acknowledgments}
PEJN gratefully acknowledges the hospitality of the
Harvard-Smithsonian Center for Astrophysics.  BRM acknowledges
grant NAS8-39073. 

\section*{References}

\ref{Baum, S.A., Heckman, T., Bridle, A., van Breugel, W., \& Miley,
G., 1988, ApJS, 68, 643}

\ref{B\"ohringer, H., Voges, W., Fabian, A.C., Edge, A.C., \& Neumann,
D.M. 1993, MNRAS, 264, L25}

\ref{Carilli, C.L., Perley, R.A., \& Harris, D.E. 1994, MNRAS, 270, 173}

\ref{David, L.P., Arnaud, K.A., Forman, W., \& Jones, C. 1990, ApJ, 356,
32}

\ref{Dwarakanath, K.S., Owen, F.N., \& van Gorkom, J.H., 1995, ApJ, 442, L1}

\ref{Ekers, R.D., \& Simkin, S.M. 1983, ApJ, 265, 85}

\ref{Hansen, L., J{\o}rgenson, H.E., \& N{\o}rgaard-Nielson, H.U. 1995, AA,
297, 13}


\ref{Heinz, S., Reynolds, C.S., \& Begelman, M.C. 1998, ApJ, 501, 126}

\ref{Ikebe, Y., Makishima, K., Ezawa, H., Fukazawa, Y., Hirayama, M., Honda, H.,
Ishisaki, Y., Kikuchi, K., Kubo, H., Murakami, T., Ohashi, T.,
Takahashi, T., \& Yamashita, K. 1997, ApJ, 481, 660}

\ref{McNamara, B.R. 1995, ApJ, 443, 77}

\ref{Melnick, J., Gopal-Krishna, \& Terlevich, R. 1997, AA, 318, 337}

\ref{Peres, C. B., Fabian, A. C., Edge, A. C., Allen, S. W., Johnstone, R. M.,
\& White, D. A. 1998, MNRAS, 298, 416}

\ref{Taylor, G.B. 1996, ApJ, 470, 394}


\ref{Taylor, G.B., Perley, R.A., Inoue, M., Kato, T., Tabara, H., \& Aizu, K.,
1990, ApJ, 360, 41}

\ref{White, D. A., Jones, C., \& Forman, W. 1997, MNRAS, 292, 419}

\vskip 0.9in

\end{document}